%% file: main.tex
\documentclass[12pt]{article}

\usepackage{scicite}

\usepackage{times}

\usepackage[version=3]{mhchem} 
\usepackage{makecell}
\usepackage{xcolor}
\usepackage{soul}
\usepackage{graphicx}
\usepackage{url}
\usepackage{float}

\newenvironment{sciabstract}{%
\begin{quote} \bf}
{\end{quote}}

\newcounter{lastnote}

\title{Large Language Model-Guided Prediction Toward Quantum Materials Synthesis}

\input{sections/authors}

\date{}

\begin{document} 


\baselineskip24pt


\maketitle


\begin{sciabstract}
\input{sections/main_text/m_01_abstract}
\end{sciabstract}


\input{sections/main_text/m_02_introduction}

\input{sections/main_text/m_03_result_discussion}

\input{sections/main_text/m_04_conclusion}

\input{sections/main_text/m_05_method}

\section*{Code Availability} 
 The source code is available at (\url{https://github.com/RyotaroOKabe/llm4syn}).

\section*{Acknowledgements}
\input{sections/main_text/m_06_acknowledge}


\bibliography{main}

\bibliographystyle{Science}

\end{document}


\beginsupplement
\tableofcontents

\input{sections/si_text/si_dataset}

\input{sections/si_text/si_training}

\input{sections/si_text/si_inference_tab}

\bibliography{main}

%% file: sections/authors.tex
\author
{Ryotaro Okabe,$^{1,2,\ast,\dagger}$ Zack West,$^{1,3,\dagger}$ Abhijatmedhi Chotrattanapituk,$^{1,3}$  \\
Mouyang Cheng,$^{1,4,5}$ Denisse Córdova Carrizales,$^{1,6}$ Weiwei Xie,$^{7}$  \\
Robert J. Cava,$^{8}$ Mingda Li$^{1,4,6,\ast}$\\
\\
\normalsize{$^{1}$Quantum Measurement Group, MIT, Cambridge, MA, USA}\\
\normalsize{$^{2}$Department of Chemistry, MIT, Cambridge, MA, USA}\\
\normalsize{$^{3}$Department of Electrical Engineering and Computer Science, MIT, Cambridge, MA, USA}\\
\normalsize{$^{4}$Center for Computational Science and Engineering, MIT, Cambridge, MA, USA}\\
\normalsize{$^{5}$Department of Materials Science and Engineering, MIT, Cambridge, MA, USA}\\
\normalsize{$^{6}$Department of Nuclear Science and Engineering, MIT, Cambridge, MA, USA}\\
\normalsize{$^{7}$Department of Chemistry, Michigan State University, East Lansing, MI, USA}\\
\normalsize{$^{8}$Department of Chemistry, Princeton University, Princeton, NJ, USA}\\
\\
\normalsize{$^{\ast}$To whom correspondence should be addressed; E-mail: rokabe@mit.edu, mingda@mit.edu.}\\
\normalsize{$^{\dagger}$These authors contributed equally to this work.}
}

%% file: sections/main_text/m_01_abstract.tex
The synthesis of inorganic crystalline materials is essential for modern technology, especially in quantum materials development. However, designing efficient synthesis workflows remains a significant challenge due to the precise experimental conditions and extensive trial and error. Here, we present a framework using large language models (LLMs) to predict synthesis pathways for inorganic materials, including quantum materials. Our framework contains three models: LHS2RHS, predicting products from reactants; RHS2LHS, predicting reactants from products; and TGT2CEQ, generating full chemical equations for target compounds. Fine-tuned on a text-mined synthesis database, our model raises accuracy from under 40\% with pretrained models, to under 80\% using conventional fine-tuning, and further to around 90\% with our proposed generalized Tanimoto similarity, while maintaining robust to additional synthesis steps. Our model further demonstrates comparable performance across materials with varying degrees of quantumness quantified using quantum weight, indicating that LLMs offer a powerful tool to predict balanced chemical equations for quantum materials discovery.

%% file: sections/main_text/m_02_introduction.tex
\section*{Introduction}
The synthesis of inorganic crystalline materials plays a critical role in modern technology, with applications ranging from energy harvesting such as solar cells\cite{Tress_SolarCellSynthesis2019} and thermoelectrics\cite{ZhouTE2018}, to advancements in quantum materials\cite{Samarth2017} and catalysts\cite{LiCatalyst2024}. Quantum materials, in particular, hold the potential to revolutionize fields such as semiconductors, quantum computing, and medical imaging, but their discovery and development are hindered by the challenges of designing synthesis experiments. The experimental conditions required for successful synthesis are often strict, involving highly controlled temperatures, pressures, and reactant purities, making trial-and-error a common approach in the experimental design\cite{schubert2019synthesis, paglione2021fundamentals}. 

Given the complexity of the synthesis processes, there is a growing demand for predictive tools that can assist in designing synthesis workflows to reduce the number of trial-and-error experiments. Machine learning (ML) has recently emerged as a powerful technique in materials science and chemistry, providing ways to automate and enhance the process of discovering new materials\cite{wei2019machine, ge2020deep}. For example, convolutional neural networks (CNNs) are used in image and spectroscopic analysis to identify structural or physical features of materials from experimental data\cite{masubuchi2020deep,Andrejevic2022XAS,cheng2024machine}. In parallel, graph neural networks (GNNs) can encode atomic structures through graph representations to predict complex materials properties\cite{xie2018crystal, okabe2023virtual}. Additionally, generative models have also been utilized to generate potential new material structures\cite{xie2021crystal, jiao2024crystal, zeni2024mattergen, merchant2023scaling, okabe2024structural}.

Despite these advances, the prediction of chemical reactions and synthesis pathways in materials remains under-explored. While previous ML models are effective at generating stable structures or predicting material properties, there has been a lack of robust techniques to address the entire experimental workflow, particularly the sequence of chemical reactions required to synthesize a given material. The ability to predict the synthesis route is critical to accelerate the experimental process and support materials scientists and solid-state chemists toward viable synthetic routes that have not been explored.

With the advent of large language models (LLMs), a new opportunity arises to leverage the capability of these models to learn from extensive, unstructured text data\cite{vaswani_attention, lagler2013gpt2, devlin2018bert, touvron2023llama, brown2020language}. Initially developed for natural language processing (NLP) tasks, LLMs have proven versatile across various domains due to their ability to capture intricate text patterns\cite{zhao2023survey}. These models can be fine-tuned on domain-specific corpora, enabling them to predict text in ways that mirror expert decision-making processes. In materials science and chemistry, research has emerged to, for example, use NLP to retrieve text data from the literature and LLMs to represent crystal structures as text \cite{chen2024mattergpt, gruver2024fine, krallinger2017information, hawizy2011chemicaltagger,schilling2024text,fu2023material}. Recent works successfully predicted pathways for chemical synthesis based on LLMs \cite{li2024transforming,kim2024large,na2023artificial, kim2024explainable}, but they are either tailored for specific material types\cite{na2023artificial,li2024transforming} or limited to simpler classification tasks\cite{kim2024large, kim2024explainable}.

In this study, we introduce an LLM-based framework capable of predicting synthesis pathways for general inorganic crystalline materials with direct applications to quantum materials synthesis. We fine-tune the models to predict chemical equations by training LLMs on a text-mined synthesis method database\cite{kononova2019text}. Three distinct models are developed: the LHS2RHS model, which predicts the products given the reactants; the RHS2LHS model, which predicts the reactants required to obtain specific products; and the TGT2CEQ model, which predicts both the left-hand and right-hand sides of a chemical equation based solely on the target compound. These models are designed to aid a materials scientist or chemist's intuition in selecting reactions and conditions with the potential to predict synthesis pathways for materials, including quantum materials whose synthesis methods are not yet established. Furthermore, we ensure the robustness of these predictions by incorporating additional prompts that describe synthesis operations, such as heating, mixing, or quenching. The models maintain consistency in predicting the chemical equations even with these additional prompts, which could negatively perturb the accuracy of the model.

We adopt a two-pronged approach for accuracy evaluation. First, we apply conventional Jaccard similarity (JS)\cite{niwattanakul2013using}  to evaluate the accuracy of the generated chemical equations. Second, we propose a new metric called the generalized Tanimoto similarity (GTS), which extends the traditional Tanimoto similarity\cite{holliday2003analysis} for individual molecules to allow the permutation of elements within chemical formulas. This new similarity improves the prediction accuracy compared to JS and provides greater flexibility in predicting chemical reactions. Our results demonstrate that the LLM-based models can predict synthesis pathways with high accuracy and flexibility, showing particular strength in generating balanced chemical equations. The LHS2RHS model achieves significant accuracy in predicting reaction products while the RHS2LHS model shows an even higher success rate in predicting the necessary reactants. 
The TGT2CEQ model can further accurately predict reactants and products based on a target product compound, making it a valuable tool for novel material discovery.
Furthermore, using the concept of quantum weight\cite{onishi2024quantum}, a recently proposed quantitative measure of a material's ``quantumness'', we assess the TGT2CEQ model's accuracy in predicting synthesis pathways for materials with varied quantumness. To our knowledge, this is the first time that LLM-based models have been used to predict the synthesis pathways of quantum materials. Our results show that materials with higher quantum weight achieve comparable or slightly higher prediction accuracy. By streamlining the design of synthesis workflows, these LLM-based models have the potential to accelerate quantum materials discovery and help researchers navigate the complexities of chemical synthesis more efficiently.

%% file: sections/main_text/m_03_result_discussion.tex
\section*{Results}

\begin{figure}[ht!]
\includegraphics[width=\textwidth]{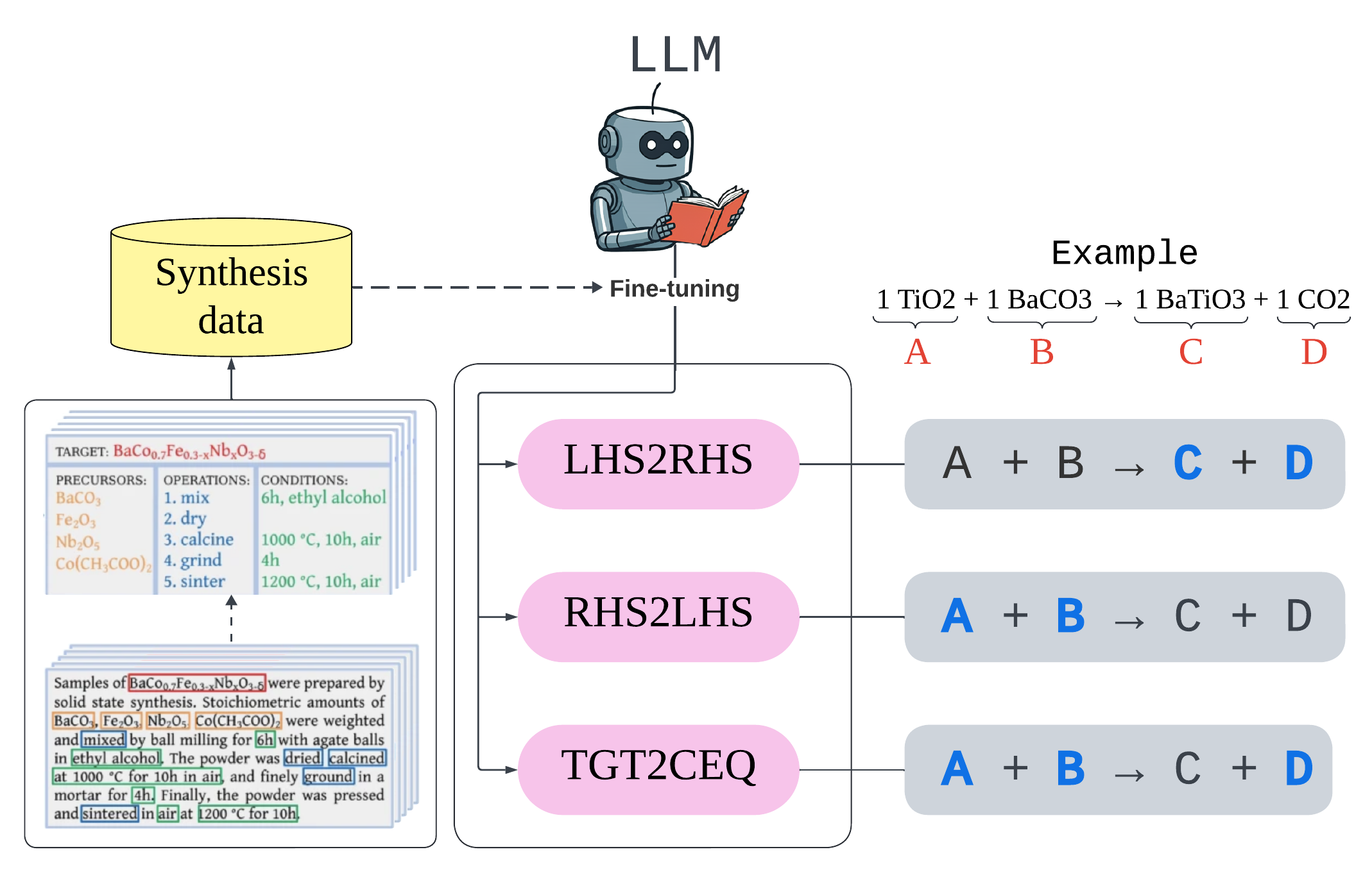}
\caption{\textbf{Overview of large language model prediction of the synthesis process.}
Synthesis protocols are pre-processed into structured data \cite{kononova2019text} (the lower left figure is taken from the same reference), which specify target compounds, precursors, operations, and conditions. The synthesis pathway of BaTiO$_3$, a ferroelectric material, is shown as an example.
The data is used to fine-tune large language models with three different models: (1) Left-hand-side-to-right-hand-side (\textbf{LHS2RHS}), where the model predicts products ($\rm{C + D}$) from given reactants ($\rm{A + B}$); (2) Right-hand-side-to-left-hand-side (\textbf{RHS2LHS}), predicting reactants from known products; and (3) Target-to-chemical-equation (\textbf{TGT2CEQ}), generating the full chemical equation from solely the target compound C. In the chemical equations, the black colors denote the input in each model, while the blue colors denote the output.}
\label{fig_overview}
\end{figure}

Figure \ref{fig_overview} shows the models that predict chemical equations and synthesis operations from known incomplete partitions. This process involves several steps: preparing the data from a text-mined synthesis method database, loading the pre-trained GPT model from HuggingFace, fine-tuning the model using the inorganic crystal synthesis database, and evaluating the model's performance. We develop three types of models: LHS2RHS, which predicts products on the right-hand side (RHS) given the reactants on the left-hand side (LHS); RHS2LHS, which predicts reactants given the products; and TGT2CEQ, which predicts the entire chemical equation (CEQ) given a target compound (TGT). We also assess the model's robustness by adding text from the synthesis operation (OPE) text to the prompts.

\subsection*{Generalized Taminoto Similarity}

Evaluating predicted text is critical for validating the performance of the fine-tuned LLM models. While standard evaluation metrics are used in NLP tasks, proper metrics must be selected to meet the demands of language generation tasks\cite{yujian2007normalized, li2013distance, papineni2002bleu, lin2004rouge}. In some situations, the exact match of the word choice and the order matters, but in other cases, the context of sentences or paragraphs is valued more than the precise correspondence of individual terms. Our work, which aims to predict chemical equations directly, requires specialized evaluation metrics different from ordinary text generation tasks to capture the structures in a chemical equation.

Chemical equations are unique because they contain chemical symbols, numerical values, and special characters. Each equation has the LHS and RHS connected by the ``$\rightarrow$'' symbol with the LHS and RHS composed of chemical formulas linked by ``+'' symbols. The order of elements in a formula and the arrangement of formulas on each side of the equation do not affect the chemical equation. For instance, ``BaTiO$_3$'', ``O$_3$TiBa'', and ``TiBaO$_3$'' are treated as the same compound, incurring no accuracy penalty in the LLM prediction. Similarly, ``BaTiO$_3$ + CO$_2$" and ``CO$_2$ + BaTiO$_3$" shall be recognized as the same equation, but should be considered different from ``TiO$_2$ + BaCO$_3$" . Furthermore, when evaluating full equations, the LHS and RHS must not mix, meaning that the reactants must not appear in the products, and vice versa. To address the unique structures in a chemical equation, we propose the GTS  as the accuracy metric.

Tanimoto similarity is widely used to evaluate the similarity between two chemical formulas, particularly for comparing the conformation of organic molecules\cite{bajusz2015tanimoto, holliday2003analysis, godden2000combinatorial}. The similarity between two chemical formulas can be computed by treating each formula as a vector in the space of elemental counts. Here, we generalize Tanimoto similarity to compare two entire chemical equations, ensuring the permutation invariance of the composition orders within each side of the reaction.

First, we represent the elemental composition of any chemical formula by a vector of element counts. Let $e_1$, $e_2$, $\dots$ be the list of elements that can appear in a chemical equation. Consider a chemical formula $c$ that, for each $i\in \{1, 2, \dots\}$, contains $n^c_{i}$ number of element $e_i$. Then, the vector representations of $c$ can be written as a vector $\mathbf{v}_c$, where each component corresponds to the count of a particular element in the formula.

\begin{equation}
    \mathbf{v}_c = [n^c_{1}, n^c_{2}, \dots].
\end{equation}

The original Tanimoto similarity between two chemical formulas, $c_1$ and $c_2$, is calculated as the cosine similarity between the vector representations of the formulas,

\begin{equation}
T(c_1, c_2) = \dfrac{\mathbf{v}_{c_1} \cdot \mathbf{v}_{c_2}}{| \mathbf{v}_{c_1} \cdot \mathbf{v}_{c_2} |}.
\end{equation}

This ensures that the similarity is normalized between 0 (no similarity) and 1 (identical composition). If the two compositions share no common elements, the similarity is zero. Otherwise, the value reflects the overlap in terms of elemental compositions.

Since Tanimoto similarity is invariant to the permutation of elements, it is not immediately suitable for comparing chemical equations due to their internal structure, which separates the LHS from the RHS. 
Moreover, each side of the equation is also sub-structured into individual reactants (from LHS) or products (from RHS), respectively. 
To address this inherent hierarchical structure in a chemical equation, we generalize the Tanimoto similarity by the following procedure:  

First, consider two sets $\mathbf{S}_1$, and $\mathbf{S}_2$ containing $|\mathbf{S}_1|$, and $|\mathbf{S}_2|$ chemical formulas:

\begin{equation}
    \mathbf{S}_n = \left\{c^{\mathbf{S}_n}_j|j\in \{1\dots |\mathbf{S}_n|\}\right\}.
\end{equation}

We define Tanimoto set similarity of $\mathbf{S}_1$ with respect to $\mathbf{S}_2$ as the average maximal Tanimoto similarity of chemical formulas in $\mathbf{S}_1$ among chemical formulas in $\mathbf{S}_2$:

\begin{equation}
    T_s(\mathbf{S}_1| \mathbf{S}_2) = \dfrac{1}{|\mathbf{S}_1|}\sum_{i}\max_{j}T(c^{\mathbf{S}_1}_i, c^{\mathbf{S}_2}_j).
\end{equation}

Second, we symmetrize the similarity by swapping the role of $\mathbf{S}_1$ and $\mathbf{S}_2$, yielding the Tanimoto set similarity between $\mathbf{S}_1$, $\mathbf{S}_2$ as

\begin{equation}
    T_s(\mathbf{S}_1, \mathbf{S}_2) = \dfrac{T_s(\mathbf{S}_1| \mathbf{S}_2)+T_s(\mathbf{S}_2| \mathbf{S}_1)}{2}.
\end{equation}
The division by 2 normalizes the similarity between 0 and 1.

Now consider two chemical equations, \textbf{eq}$_1$ and \textbf{eq}$_2$, where each equation contains reactants and products. Since the ordering of chemical formulas in each side of the reactions does not change the equation, we can represent reactants and products of equation \textbf{eq}$_i$ as sets $\mathbf{R}_i$ and $\mathbf{P}_i$, respectively. Specifically, each chemical equation can be represented as:

\begin{equation}
\mathbf{eq}_i: \mathbf{R}_i \rightarrow \mathbf{P}_i.
\end{equation}

After splitting the equations into reactants and products, to preserve non-mixing properties, the similarity between $\mathbf{eq}_1$, and $\mathbf{eq}_2$ depends only on the similarity between the $\mathbf{R}$'s, and the similarity between the $\mathbf{P}$'s with Tanimoto set similarity defined above. 

Finally, we define our GTS for two chemical equations $\mathbf{eq}_1$ and $\mathbf{eq}_2$ as the summation of these two Tanimoto set similarities.  

\begin{equation}
\Bar{T}(\mathbf{eq}_1, \mathbf{eq}_2) = \dfrac{T_s(\mathbf{R}_1, \mathbf{R}_2)+T_s(\mathbf{P}_1, \mathbf{P}_2)}{2}
\end{equation}
The division by 2 is for normalization of similarity between 0 and 1.

\begin{figure}[ht!]
\begin{center}
\includegraphics[width=0.8\textwidth]{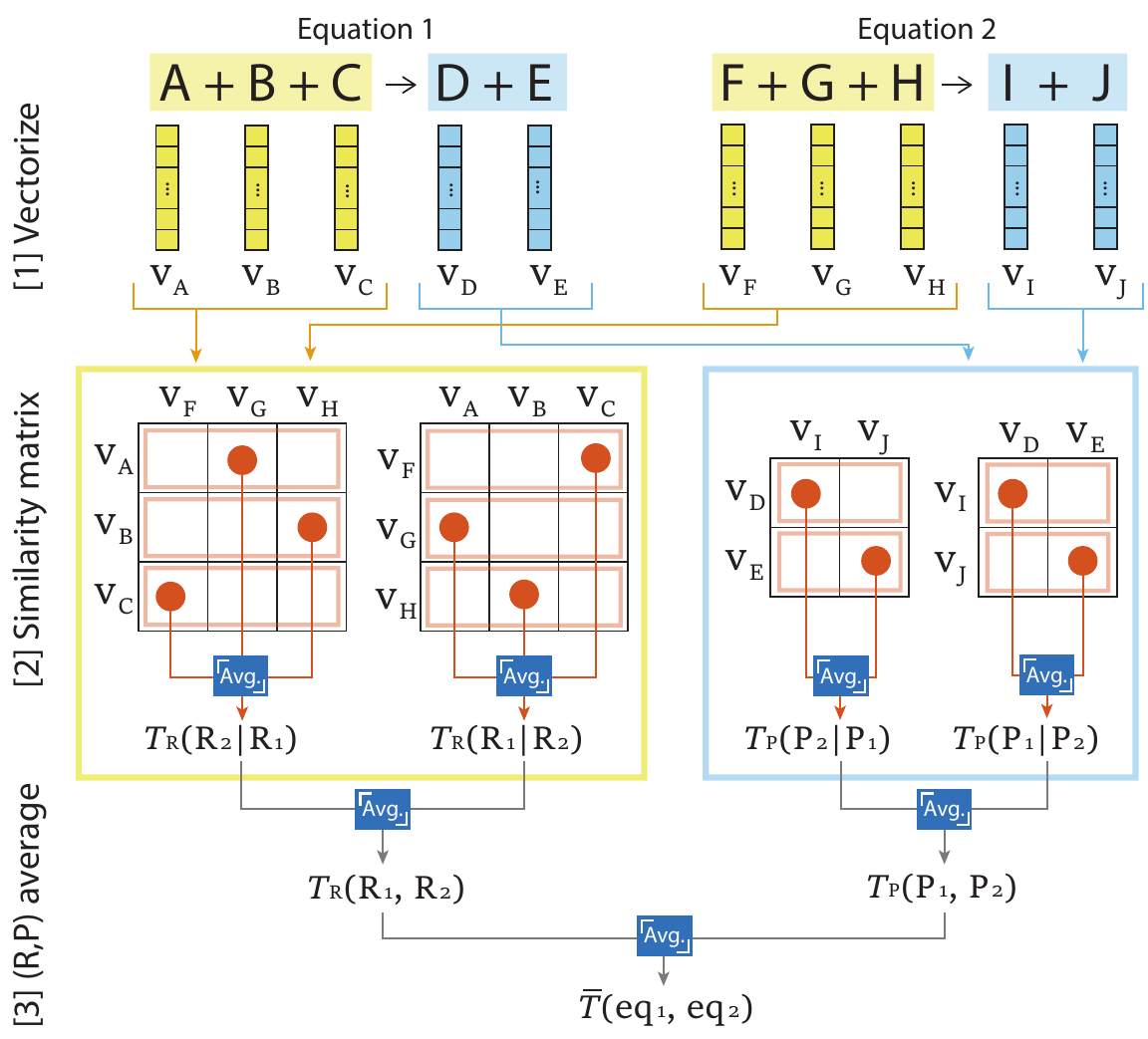}
\caption{\textbf{Workflow to compute the generalized Tanimoto similarity (GTS).} For any pair of predicted and ground truth chemical equations, each chemical formula, represented by A, B, ... J, from both equations is vectorized into vector of element counts $\mathbf{v}_\text{A},\mathbf{v}_\text{B},...\mathbf{v}_\text{J}$ and partitioned into reactants (yellow), and products (blue) sets. We calculate the Tanimoto similarity between every ordered pair of vectors from Equations 1 and 2, resulting in similarity matrices for reactants and products. Then, we extract Tanimoto set similarities ($T_s(\mathbf{R}_1,\mathbf{R}_2)$, $T_s(\mathbf{P}_1,\mathbf{P}_2)$) by iterating through rows of both matrices, selecting the maximum Tanimoto similarity for each row and averaging them ($T_s(\mathbf{R}_1|\mathbf{R}_2)$, $T_s(\mathbf{P}_1|\mathbf{P}_2)$). This effectively pairs each chemical formula from Equation 1 to a chemical formula from Equation 2 that is most similar. We then symmetrize the similarities by applying the same procedure as above, switching the roles of Equation 1 and Equation 2 ($T_s(\mathbf{R}_2|\mathbf{R}_1)$, $T_s(\mathbf{P}_2|\mathbf{P}_1)$). Finally, the GTS $\bar{T}(\mathbf{eq}_1, \mathbf{eq}_2)$ is calculated by averaging the Tanimoto set similarities of reactants and products.}
\label{fig_similaritymethod}
\end{center}
\end{figure}

\subsection*{Prediction of one side of a chemical equation from the other (LHS2RHS and RHS2LHS)}

In this study, we fine-tune LLMs to predict the reactants of a chemical equation given the products or vice versa. This ability is essential for chemists to predict the outcome of a reaction from given reactants (LHS2RHS) or reverse-engineer the reactants needed for a specific target product (RHS2LHS). We also expand the model's capability to predict entire chemical equations given target compounds (TGT2CEQ). Our fine-tuned LLMs enable more efficient experimental workflows and reduce trial and error. 
Additional prediction results are provided in the SI Tables S1 through S4.

\subsubsection*{LHS2RHS and RHS2LHS model performance}

\begin{figure}[]
\includegraphics[width=\textwidth]{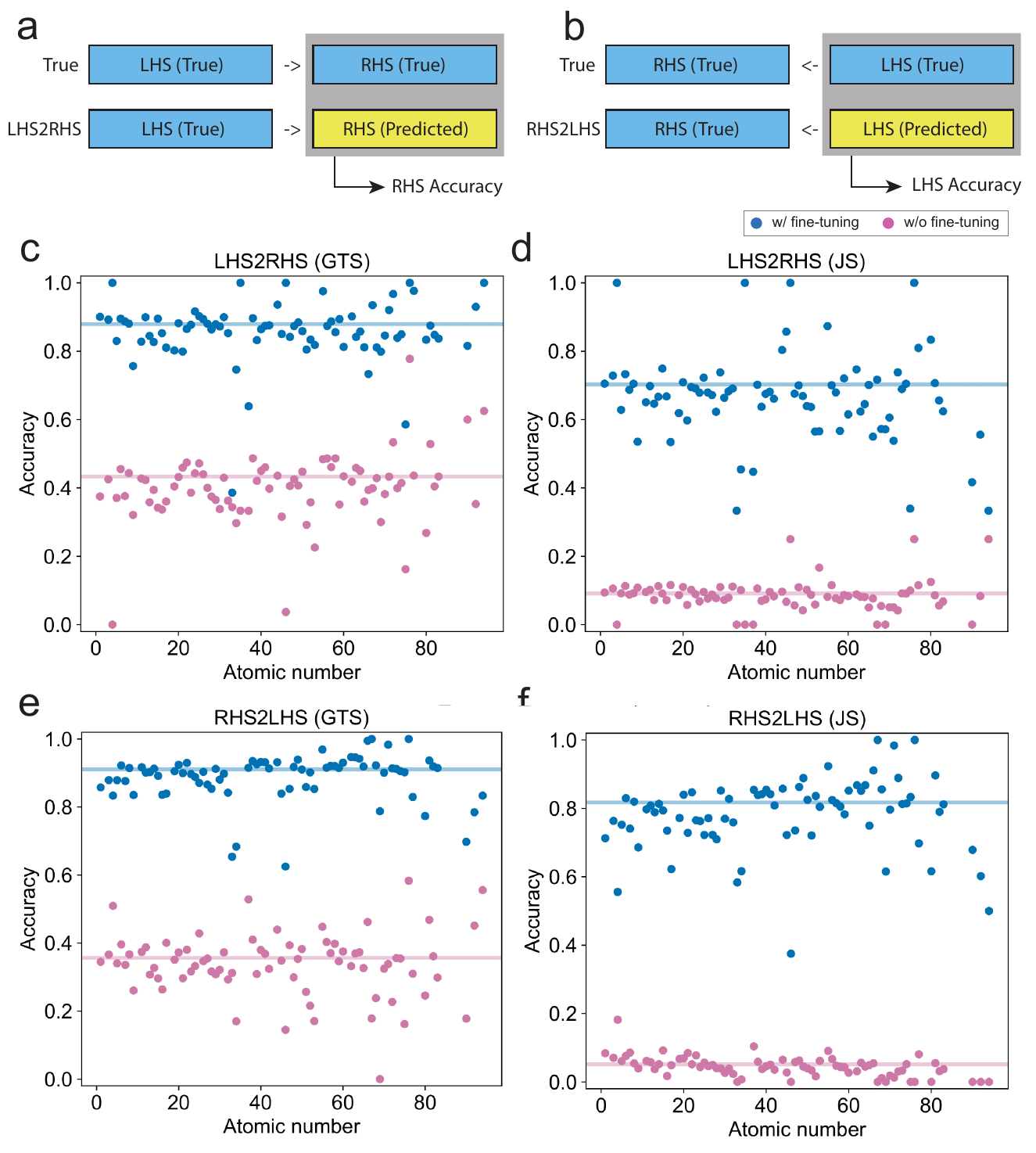}
\caption{\textbf{Schematics and average prediction accuracies for LHS2RHS and RHS2LHS.} (a) LHS2RHS schematic, where reactants are given as input and products are generated. (b) RHS2LHS schematic, where products are given as input and reactants are generated. (c-d) GTS and JS for LHS2RHS predictions. (e-f) GTS and JS for RHS2LHS predictions. Models with (w/) fine-tuning (blue) perform significantly better than models (pink) without (w/o) fine-tuning. Our proposed GTS shows a notably higher accuracy than conventional JS accuracy.}
\label{fig_lhs_rhs}
\end{figure}

The LHS2RHS model predicts the products based on the given reactants. The RHS2LHS model, in contrast, predicts the necessary reactants for target products. The schematics for these tasks are illustrated in Figures \ref{fig_lhs_rhs}a-b.

To evaluate the model’s performance, we use JS as a benchmark and show our proposed GTS can lead to performance improvement. These metrics assess the accuracy of predicted chemical formulas relative to the ground truth. The GTS is more tolerant to variations, such as the permutation of chemical elements within the formula. The results, shown in Figures \ref{fig_lhs_rhs}c-f, demonstrate the improved prediction accuracy after fine-tuning the models with the CEDER dataset. We plot average prediction accuracy against the atomic number of elements in the equation. The fine-tuned model achieves a GTS of 0.879 for LHS2RHS and 0.911 for RHS2LHS. JS shows lower performance (0.702 and 0.817, respectively) even after fine-tuning than GTS, demonstrating the power of our proposed GTS to capture the similarity of the chemical formulas under the acceptable variances of symbol orders.

In contrast, the pre-trained LLM without fine-tuning generates repetitive or irrelevant sequences, resulting in lower accuracy (shown in pink in Figures \ref{fig_lhs_rhs}c-f). This highlights the importance of fine-tuning for capturing the chemical relationships between reactants and products.

Tables \ref{tab_main_lhs2rhs} and \ref{tab_main_rhs2lhs} provide a few examples of the LHS2RHS and RHS2LHS model predictions. In these examples, the fine-tuned models perform well, often producing outputs that closely match the ground truth. However, as observed, RHS2LHS typically achieves higher accuracy, likely due to the relatively limited diversity of reactants compared to the larger space of possible products in material synthesis. 

\begin{table}[h!]
\begin{center}
\caption{Selected LHS2RHS prediction results using transition metal and rare-earth oxides with relatively high quantum weight as examples. The LLM predicts the chemical products (RHS) as output from the chemical reactants (LHS) as input prompt. The right arrow ``$\rightarrow$" serves as a separation delimiter for the LLM to denote the LHS of a chemical equation as input. Ground truths and LLM predictions with and without fine-tuning are compared. }
\label{tab_main_lhs2rhs}
\begin{tabular}{c|p{5cm}|p{6.5cm}|c|c}
    Model & Prompt (LHS) & Output (RHS) & GTS & JS \\
    \Xhline{5\arrayrulewidth}
    Ground truth & 2 MnO$_2$ + 0.5 Li$_2$CO$_3$ $\rightarrow$ & 1 LiMn$_2$O$_4$ + 0.5 CO$_2$ + 0.25 O$_2$ & 1.0 & 1.0 \\  
    \hline
    w/ fine-tuning & 2 MnO$_2$ + 0.5 Li$_2$CO$_3$ $\rightarrow$ & 1 LiMn$_2$O$_4$ + 0.5 CO$_2$ + 0.25 O$_2$ & 1.0 & 1.0 \\  
    \hline
    w/o fine-tuning & 2 MnO$_2$ + 0.5 Li$_2$CO$_3$ $\rightarrow$ & 0.5 Li$_2$CO$_3$ $\rightarrow$ 0.5 Li$_2$CO$_3$ $\rightarrow$ 0.5 & 0.465 & 0.091  \\  
    \Xhline{5\arrayrulewidth}
    Ground truth & 0.25 Na$_2$CO$_3$ + 0.005 MnO + 0.497 Nb$_2$O$_5$ + 0.25 K$_2$CO$_3$ + 0.004 O$_2$ $\rightarrow$ & 1 (Na$_{0.5}$K$_{0.5}$)(Nb$_{0.995}$Mn$_{0.005}$)O$_3$ + 0.5 CO$_2$ & 1.0 & 1.0 \\  
    \hline
    w/ fine-tuning & 0.25 Na$_2$CO$_3$ + 0.005 MnO + 0.497 Nb$_2$O$_5$ + 0.25 K$_2$CO$_3$ + 0.004 O$_2$ $\rightarrow$ & 1 (K$_{0.5}$Na$_{0.5}$)(Nb$_{0.995}$Mn$_{0.005}$)O$_3$ + 0.5 CO$_2$ & 1.0 & 0.667 \\  
    \hline
    w/o fine-tuning & 0.25 Na$_2$CO$_3$ + 0.005 MnO + 0.497 Nb$_2$O$_5$ + 0.25 K$_2$CO$_3$ + 0.004 O$_2$ $\rightarrow$ & 0.001 & 0.0 & 0.0 \\  
    \Xhline{5\arrayrulewidth}
    Ground truth & 3 TiO$_2$ + 0.5 Fe$_2$O$_3$ + 1.5 La$_2$O$_3$ + 1 Bi$_2$O$_3$ $\rightarrow$ & 1 Bi$_2$La$_3$Ti$_3$FeO$_{15}$ & 1.0 & 1.0 \\  
    \hline
    w/ fine-tuning & 3 TiO$_2$ + 0.5 Fe$_2$O$_3$ + 1.5 La$_2$O$_3$ + 1 Bi$_2$O$_3$ $\rightarrow$ & 1 Bi$_3$La$_3$Ti$_3$FeO$_{15}$ & 0.996 & 0.333 \\  
    \hline
    w/o fine-tuning & 3 TiO$_2$ + 0.5 Fe$_2$O$_3$ + 1.5 La$_2$O$_3$ + 1 Bi$_2$O$_3$ $\rightarrow$ & 0.5 Fe$_2$O$_4$ + 0.5 & 0.226 & 0.0 \\  
    \Xhline{5\arrayrulewidth}
\end{tabular}
\end{center}
\end{table}

\begin{table}[h!]
\begin{center}
\caption{Selected RHS2LHS results, using transition metal and rare-earth oxides as examples. The LLM predicts the chemical reactants (LHS) as output from the chemical products (RHS) as input prompt. The left arrow ``$\leftarrow$" serves as a separation delimiter for LLM to denote the RHS of a chemical equation as input. Ground truths and LLM predictions with (w/) and without (w/o) fine-tuning are compared. }
\label{tab_main_rhs2lhs}
\begin{tabular}{c|p{5cm}|p{6.5cm}|c|c}
    Model & Prompt (RHS) & Output (LHS) & GTS & JS \\
    \Xhline{5\arrayrulewidth}
    Ground truth &  1 CaCu$_3$Ti$_4$O$_{12}$ + 1 CO$_2$ $\leftarrow$ & 3 CuO + 4 TiO$_2$ + 1 CaCO$_3$  & 1.0 & 1.0  \\  
    \hline
    w/ fine-tuning & 1 CaCu$_3$Ti$_4$O$_{12}$ + 1 CO$_2$ $\leftarrow$ & 3 CuO + 4 TiO$_2$ + 1 CaCO$_3$ & 1.0 & 1.0 \\  
    \hline
    w/o fine-tuning & 1 CaCu$_3$Ti$_4$O$_{12}$ + 1 CO$_2$ $\leftarrow$ & 1 CO$_2$ $\leftarrow$ 1 CO$_2$ $\leftarrow$ 1 CO$_2$ $<$ & 0.519 & 0.0 \\  
    \Xhline{5\arrayrulewidth}
    Ground truth & 1 LiMn$_2$O$_4$ + 1.5 H$_2$O + 0.25 O$_2$ $\leftarrow$ & 1 LiOH·H$_2$O + 2 $\beta$-MnO$_2$  & 1.0 & 1.0  \\  
    \hline
    w/ fine-tuning & 1 LiMn$_2$O$_4$ + 1.5 H$_2$O + 0.25 O$_2$ $\leftarrow$ & 1 LiOH·H$_2$O + 2 MnO$_2$ x & 1.0 & 0.571 \\  
    \hline
    w/o fine-tuning & 1 LiMn$_2$O$_4$ + 1.5 H$_2$O + 0.25 O$_2$ $\leftarrow$ & 1.5 O$_2$ $\leftarrow$ 1.5 O$_2$ $\leftarrow$ 1. & 0.6 & 0.0 \\  
    \Xhline{5\arrayrulewidth}
    Ground truth &  1 TbBaMn$_2$O$_6$ + 3 CO$_2$ $\leftarrow$ & 1 BaCO$_3$ + 2 MnCO$_3$ + 0.25 Tb$_4$O$_7$ + 0.625 O$_2$  & 1.0 & 1.0 \\
    \hline
    w/ fine-tuning &  1 TbBaMn$_2$O$_6$ + 3 CO$_2$ $\leftarrow$ & 1 Mn$_2$O$_3$ + 1 BaCO$_3$ + 0.25 Tb$_4$O$_7$ + 0.125 O$_2$  & 0.962 & 0.545 \\
    \hline
    w/o fine-tuning &  1 TbBaMn$_2$O$_6$ + 3 CO$_2$ $\leftarrow$ & TbBaMn$_2$O$_6$ + 3 CO$_2$ $\leftarrow$ TbBaMn$_2$O$_6$ + 3 CO$_2$ $\leftarrow$  & 0.548 & 0.077 \\
    \Xhline{5\arrayrulewidth}
\end{tabular}
\end{center}
\end{table}

\subsection*{Predicting whole chemical equation given target compound (TGT2CEQ)}

Building on the success of LHS2RHS and RHS2LHS, we now demonstrate the TGT2CEQ model, which predicts both reactants and products given only a target compound. This task is critical for chemists, who often start with a target material and must infer the entire chemical equation for its synthesis.

\begin{figure}[]
\includegraphics[width=\textwidth]{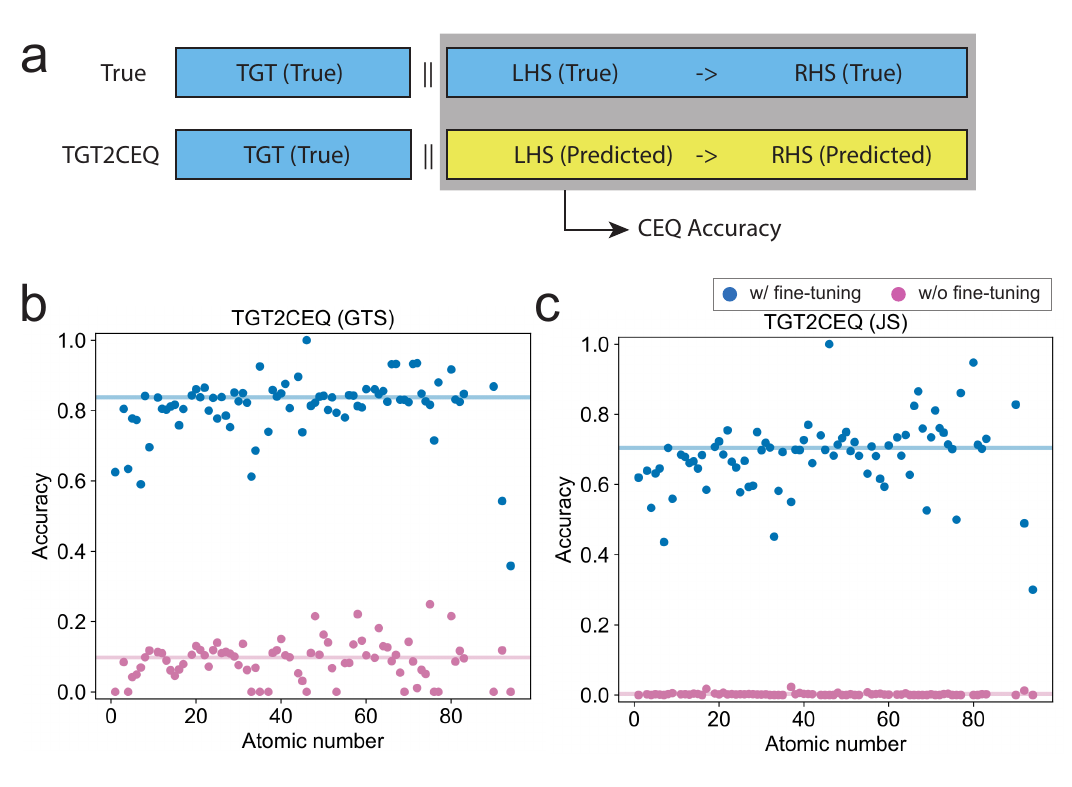}
\caption{\textbf{Schematics and average prediction accuracies for TGT2CEQ.} (a) TGT2CEQ schematic, where the target compound is given as input and the full chemical equation is generated. (b-c) GTS and JS for models (blue) with (w/) fine-tuning and models (pink) without (w/o) fine-tuning. Our proposed GTS shows a consistent higher accuracy than using conventional JS. 
}
\label{fig_tgt2ceq}
\end{figure}

Figure \ref{fig_tgt2ceq}a illustrates the setup for this prediction, where the target compound serves as the input prompt, followed by the separator symbol ``$||$.'' The model generates the entire chemical equation, predicting both reactants (LHS) and products (RHS). Using GTS, the accuracy for this task reaches 0.838 after fine-tuning, a substantial improvement over the pre-trained model’s near-zero accuracy (0.098). Also, it improves from the JS metric with an accuracy of 0.704.

Table \ref{tab_main_tgt2ceq} provides a few examples of the predicted chemical equations for target compounds. The TGT2CEQ model frequently produces physically feasible reactions, demonstrating its potential for generating novel synthetic routes, even for complex target compounds. 

\begin{table}[h!]
\begin{center}
\caption{Selected examples of TGT2CEQ results. The LLM predicts the full chemical equation (CEQ) from the target compound as input prompt (TGT). The symbol "$||$" serves as a separation delimiter for LLM to denote the target compound as the input. Ground truths and LLM predictions with (w/) and without (w/o) fine-tuning are compared. }
\label{tab_main_tgt2ceq}
\begin{tabular}{c|p{4cm}|p{8cm}|c|c}
    Model & Prompt (TGT) & Output (CEQ) & GTS & JS \\
    \Xhline{5\arrayrulewidth}
    Ground truth & CaTi$_4$Cu$_3$O$_{12}$ $||$ & 3 CuO + 4 TiO$_2$ + 1 CaCO$_3$ $\rightarrow$ 1 CaCu$_3$Ti$_4$O$_{12}$ + 1 CO$_2$ & 1.0 & 1.0 \\
    \hline
    w/ fine-tuning & CaTi$_4$Cu$_3$O$_{12}$ $||$ & 3 CuO + 4 TiO$_2$ + 1 CaCO$_3$ $\rightarrow$ 1 CaCu$_3$Ti$_4$O$_{12}$ + 1 CO$_2$ & 1.0 & 1.0 \\
    \hline
    w/o fine-tuning & CaTi$_4$Cu$_3$O$_{12}$ $||$ & $||$ $||$ $||$ $||$ $||$ $||$ $||$ $||$ $||$ $||$ $||$ $||$ $||$ $||$ $||$ $||$ & 0.0 & 0.0 \\
    \Xhline{5\arrayrulewidth}
    Ground truth & Zr$_{0.54}$Ti$_{0.46}$PbO$_3$ $||$ & 1 PbO + 0.46 TiO$_2$ + 0.54 ZrO$_2$ $\rightarrow$ 1 Pb(Zr$_{0.54}$Ti$_{0.46}$)O$_3$ & 1.0 & 1.0 \\
    \hline
    w/ fine-tuning & Zr$_{0.54}$Ti$_{0.46}$PbO$_3$ $||$ & 1 PbO + 0.46 TiO$_2$ + 0.54 ZrO$_2$ $\rightarrow$ 1 PbZr$_{0.54}$Ti$_{0.46}$O$_3$ & 1.0 & 0.8 \\
    \hline
    w/o fine-tuning & Zr$_{0.54}$Ti$_{0.46}$PbO$_3$ $||$ & 0x0 0x0 0x0 0x0 0x0 0x0 0x0 0x0 0x0 0x0 0x0 0x0 0x0 0 & 0.0 & 0.0 \\
    \Xhline{5\arrayrulewidth}
    Ground truth & Zr$_{20}$Nb$_{20}$Ni$_{60}$ $||$ & 20 Zr + 60 Ni + 20 Nb $\rightarrow$ 1 Ni$_{60}$Nb$_{20}$Zr$_{20}$ & 1.0 & 1.0 \\
    \hline
    w/ fine-tuning & Zr$_{20}$Nb$_{20}$Ni$_{60}$ $||$ & 20 Zr + 60 Ni + 20 Nb $\rightarrow$ 1 Ni$_{60}$Zr$_{20}$Nb$_{20}$ & 1.0 & 0.8 \\
    \hline
    w/o fine-tuning & Zr$_{20}$Nb$_{20}$Ni$_{60}$ $||$ & 0 $||$ 0 $||$ 0 $||$ 0 $||$ 0 $||$ 0 $||$ 0 $||$ 0 $||$ & 0.0 & 0.0 \\
    \Xhline{5\arrayrulewidth}
\end{tabular}
\end{center}
\end{table}

\subsection*{Robustness of predictions with additional prompts}

To assess the robustness of the models, we introduce additional synthesis operations (such as heating, mixing, or quenching) into the prompts. Models with these additional prompts are denoted LHSOPE2CEQ, RHSOPE2CEQ, and TGTOPE2CEQ. These variants evaluate the model's ability to handle more complex inputs that include synthesis instructions in addition to chemical components.

\begin{figure}[]
\includegraphics[width=0.9\textwidth]{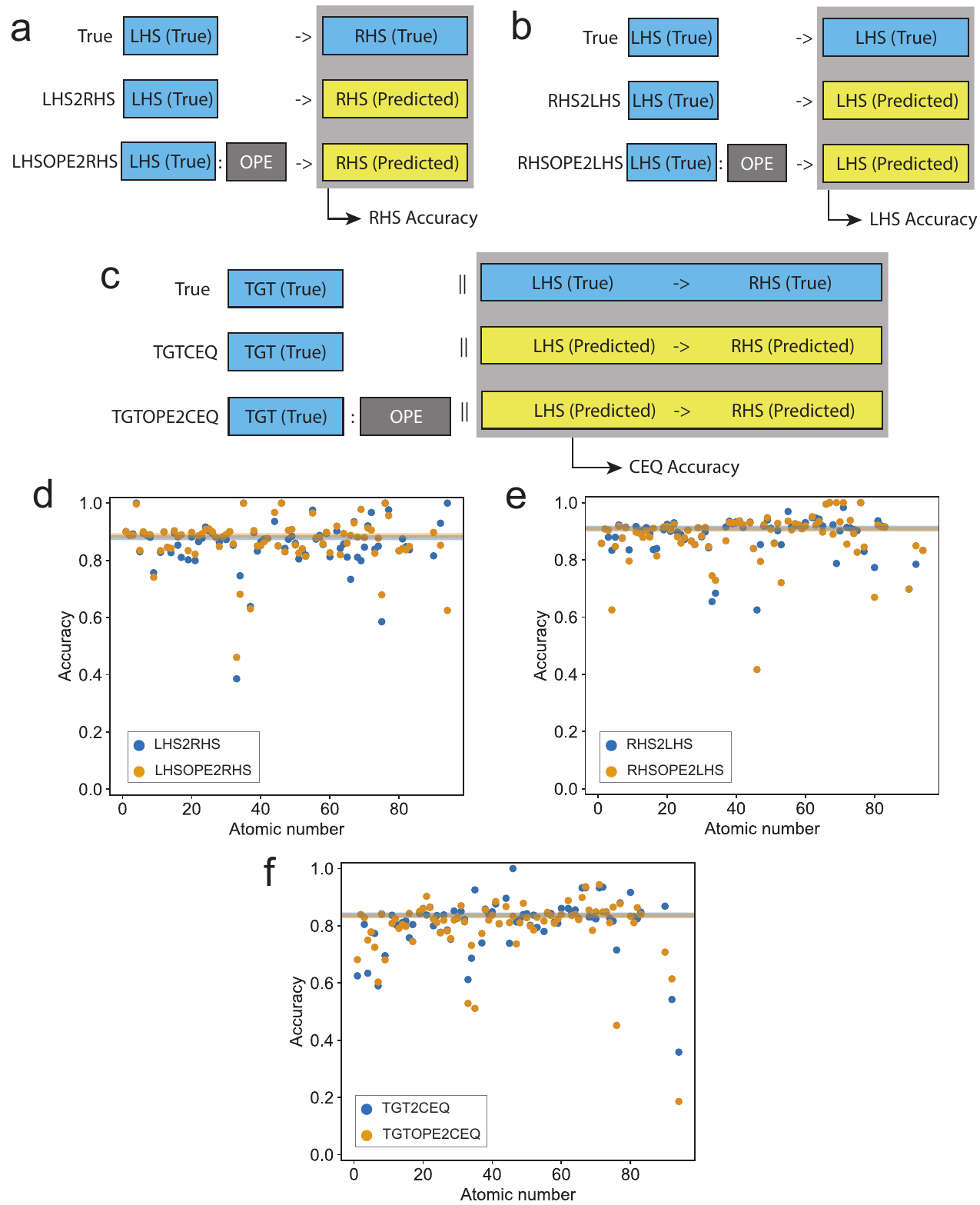}
\caption{\textbf{Robustness of chemical equation predictions with additional synthesis operation prompts.} (a-c) Representations of LHSOPE2RHS, RHSOPE2LHS, and TGTOPE2CEQ where synthesis operations are included as additional prompts. (d-f) Prediction accuracy without (blue) and with (orange) these additional prompts using GTS. The preservation of prediction accuracy shows robustness against additional operation prompts.}
\label{fig_ope}
\end{figure}

Figure \ref{fig_ope} shows that incorporating these synthesis operations into the input prompts do not significantly reduce prediction accuracy. The model's robustness is preserved with the accuracy difference remaining below 0.8\%. Table \ref{tab_main_all} summarizes the prediction accuracy for the various models with and without synthesis operation prompts. It is worthwhile mentioning that the presence of additional input prompts does not reduce the prediction accuracy, indicating our model's robustness. 

\begin{table}[h!]
\begin{center}
\caption{Prediction accuracies with and without synthesis operations (OPE).}
\label{tab_main_all}
\begin{tabular}{c|cccc}
    Metric & \makecell{GTS \\ (w/ fine-tuning)} & \makecell{JS \\ (w/ fine-tuning)} & \makecell{GTS \\ (w/o fine-tuning)} & \makecell{JS \\ (w/o fine-tuning)}  \\
\hline
    LHS2RHS & 0.879 & 0.702 & 0.434 & 0.091   \\ 
    RHS2LHS & 0.911 & 0.817 & 0.356 & 0.052   \\
    TGT2CEQ & 0.838 & 0.704 & 0.098 & 0.003  \\
    LHSOPE2RHS & 0.886 & 0.709 & 0.037 & 0.011   \\
    RHSOPE2LHS & 0.907 & 0.812 & 0.121 & 0.030   \\
    TGTOPE2CEQ & 0.835 & 0.705 & 0.017 & 0.004   \\
\end{tabular}
\end{center}
\end{table}

\begin{figure}[]
\includegraphics[width=\textwidth]{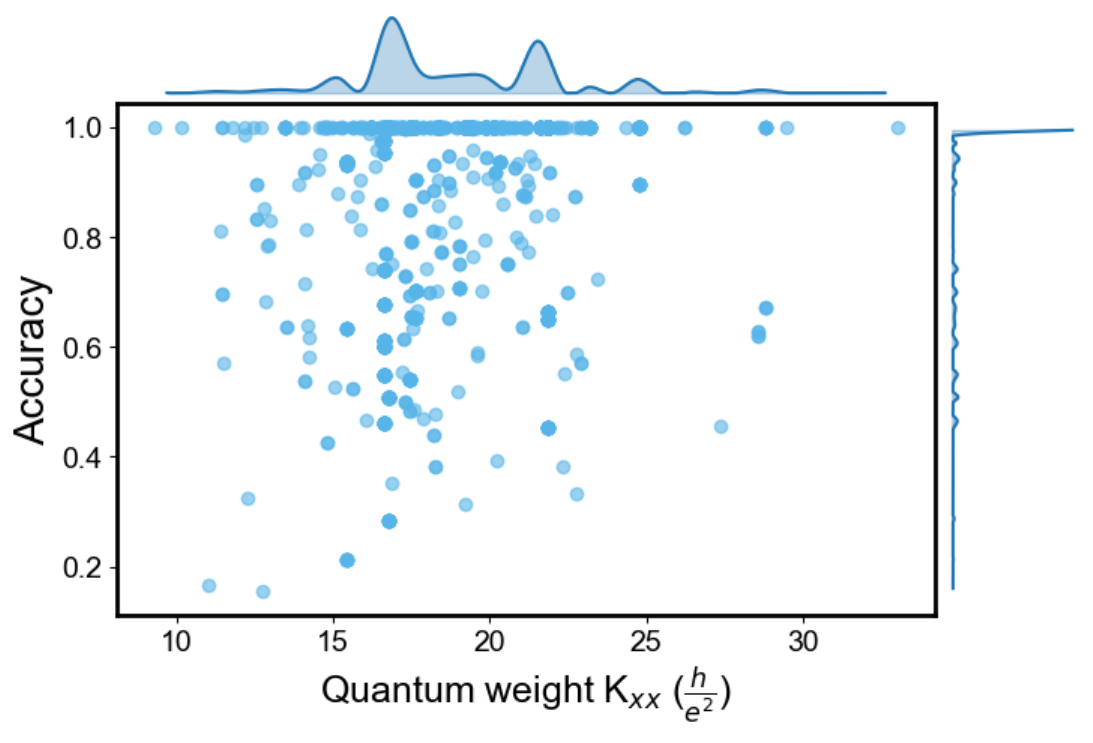}
\caption{\textbf{Scatter plot of TGT2CEQ-predicted GTS accuracy a function of quantum weight.} The curves on the top of, and on the right of the plot represent projected probability distributions of the data points along quantum weight, and GTS accuracy axes, respectively. The Pearson correlation coefficient is $r = 0.26$, indicating a weak but positive correlation between materials' quantum weight and model accuracy. This implies that the model can be applied to materials with high quantum weight (i.e., high ``quantumness'') with comparable or even higher accuracy.}
\label{fig_tgt2ceq_qw}
\end{figure}

\subsection*{Model accuracy with respect to the level of quantumness}
To understand our model's accuracy when predicting the synthesis pathway of a quantum material, we test the correlation between our model's accuracy against the quantum weight $K_{xx}$. Quantum weight is a recently proposed quantitative metric that captures the quantum geometry and topology of ground states in gapped materials\cite{onishi2024quantum}. Higher quantum weight indicates a greater degree of ``quantumness." We compute quantum weights for materials with a bandgap greater than 0.2 eV, given that quantum weight is only applicable to gapped systems, from the imaginary part of optical conductivity Im$\sigma(\omega)$, which is screened by using a pre-trained graph neural network by Hung et al.\cite{hung2024universal}. This graph neural network predicts optical properties directly from the crystal structure. The network achieves high fidelity using accurate high-throughput DFT calculations and ensemble embedding learning.

Following this procedure, we plot the predicted GTS accuracy versus quantum weight for 1,206 structures in Fig. \ref{fig_tgt2ceq_qw}. The Pearson correlation analysis of the plot yields a positive coefficient of 0.26, indicating the model performs comparably on materials with high ``quantumness.''

%% file: sections/main_text/m_04_conclusion.tex
\section*{Discussion}

Our work demonstrates the application of LLMs to predict chemical synthesis pathways for general inorganic materials, particularly quantum materials. By fine-tuning pre-trained LLMs on a text-mined synthesis method database, we develop three distinct models: LHS2RHS, RHS2LHS, and TGT2CEQ. These models enable relatively accurate predictions of chemical equations.  The LHS2RHS and RHS2LHS models predict either the products or reactants given the opposite side of a chemical equation and the TGT2CEQ model predicts the entire equation based on a target compound.

Using JS as our benchmark metric, our fine-tuned models significantly improve prediction accuracy compared to models without fine-tuning. Our proposed GTS metric further improves the prediction accuracy. 
This is so since our proposed GTS captures the hierarchical invariant structures in a chemical equation, which could improve the accuracy of the prediction. The LHS2RHS and RHS2LHS models achieve high precision--with the RHS2LHS model consistently outperforming LHS2RHS due to the relative simplicity of reactants compared to products. Furthermore, the TGT2CEQ model demonstrates its capability to generate balanced chemical equations, providing valuable insights into potential synthetic routes for materials where no existing synthesis methods have been established. A key strength of our approach is its robustness to additional synthesis operation prompts. The models maintain high accuracy even when presented with more complex prompts, showing their versatility in handling real-world synthesis data. 
Our LLM applies to quantum materials for TGT2CEQ tasks, preserving accuracy even for materials with higher quantumness. It directly predicts complete chemical equations or compounds without predefined restrictions, enabling the prediction of full synthesis recipes from target compounds. Our results also indicate that quantum materials with different level of quantumness may share similar complexities from a perspective of synthesis. 

There are several promising avenues to enhance the predictive capabilities of these models. One potential direction is to incorporate more advanced versions of GPT models, such as o1, which may offer greater depth and precision in understanding chemical reactions and predicting synthesis recipes for versatile quantum materials. Better integration of structural and compositional data to improve predictions could be enabled by developing multi-modal models that combine LLMs with other forms of machine learning, such as graph neural networks or generative models.

Another promising direction involves incorporating the intuition and expertise of synthesis experts into the model’s learning process through active learning. In this approach, the model would iteratively query human experts for feedback on uncertain or ambiguous predictions, allowing it to refine its understanding in a more targeted and efficient manner. This involvement of domain experts could help the model prioritize the most informative data and learn more effectively from fewer experiments, thereby improving the accuracy of predictions in real-world synthesis scenarios. Combining data-driven insights with expert knowledge bridges the gap between theoretical predictions and practical experimental workflows, making the models more applicable to real-world tasks, such as synthesizing quantum materials.

%% file: sections/main_text/m_05_method.tex
\section*{Materials and Methods}

\subsection*{Data preparation}
Ideally, we would use a quantum materials synthesis database for LLM training. However, there is no consensus on the definition of ``quantum materials,'' which can include quantum statistics, quantum correlation, quantum entanglement, or even the quantization of lattice degrees of freedom. Therefore, we use an inorganic materials synthesis database, the CEDER database, for LLM training. The CEDER database is a text-mined dataset of inorganic material synthesis recipes \cite{kononova2019text}. The database contains 19,488 synthesis records extracted from 53,538 paragraphs of scientific literature using natural language processing (NLP) techniques, many of which could be considered quantum materials, particularly those with transition metals or rare-earth oxides. Each record contains balanced reactions, target materials, and synthesis operations. We organize the data into prompt-target pairs for LLM training. 
Details of the dataset are shown in Supplementary Information (SI) Figure S1. 
The prompt is the user input, and the target text is the expected output. In chemical equations, an arrow ($\rightarrow$) links the LHS and RHS. A separator marks the boundary between the prompt and the target text. Notes related to target compounds or additives in the CEDER database are removed, leaving only the chemical equations.

If synthesis operations (OPE) are used as additional prompts, we simplify them to sequences of operation names while omitting details such as temperature profiles and durations. The list of operation names includes ``SolutionMix," ``Shape," ``Dry," ``Mix," ``Heat," and ``LiquidGrind." 

\subsection*{Fine-tuning the model}

The fine-tuning process begins by loading the Distilled-GPT2 (distilGPT2) model from HuggingFace as the pre-trained model\cite{lagler2013gpt2, sanh2019distilbert, wolf2019huggingface}. We utilize the HuggingFace API and the transformers library to load the model, using the class \texttt{AutoModelForCausalLM} for model loading and \texttt{AutoTokenizer} for tokenization. We further apply the\\ \texttt{DataCollatorForLanguageModeling} to manage dynamic padding during training.

The training process employs a learning rate of $2 \times 10^{-5}$, a weight decay of 0.01, and a batch size of 4. The dataset is split into training and test sets at a ratio of 9:1. We train the model for 100 epochs using 10-fold cross-validation and repeat the process twice (200 epochs) to ensure the convergence of the loss function. Details of the training process are provided in SI Figure S3.

\subsection*{Text inference}

The text inference process relies on an autoregressive model for chemical equation prediction. 
To balance the diversity and accuracy of text inference, we employ beam-search multinomial sampling with two beams, enabled sampling, and one beam group. This approach generates multiple potential sequences and selects the highest overall probability.

Autoregressive models can generate redundant text by repeating phrases given in the prompts. To mitigate this issue, we specify a fixed text length equal to the ground truth equation length, which ensures the model stops generating after producing the desired output. This approach improves the accuracy of generated chemical equations. An analysis of the relationship between text lengths for LHS, RHS, TGT, and CEQ is included in the SI Figure S2.

\subsubsection*{Jaccard Similarity (JS)}
In addition to the proposed GTS, we also employ the commonly-used JS as a benchmarking metric\cite{niwattanakul2013using}. The JS measures the similarity between two sets by comparing their intersection and union. For two sets of formulas (e.g., reactants or products), the JS is given by:

\begin{equation}
J(A, B) = \frac{|A \cap B|}{|A \cup B|}
\end{equation}

Sets $A$ and $B$ represent the formulas found in their respective chemical equations. The JS is used to measure the structural similarity between two chemical equations based on the overlap of their components. However, our GTS offers a distinct advantage: it allows for the permutation of atoms within a formula and takes a more flexible approach to evaluating chemical formulas.
Unlike JS, which requires the predicted formula to match the ground truth exactly, the GTS can still assess similarity even when the two formulas are not identical. This flexibility makes the GTS better suited for predicting chemical equations.
The original JS also evaluates the LHS and RHS of a chemical equation together. This creates the risk of assigning a similarity when components from the LHS appear on the RHS, which can be misleading. Our GTS addresses this issue and offers a more accurate similarity assessment for chemical equation prediction.

\subsubsection*{Evaluation of quantum materials using quantum weight}
To evaluate our LLM's prediction of quantum material recipes, we need a quantified descriptor to identify potential quantum material candidates from the inorganic database.
However, a general metric for evaluating the ``quantumness" of materials is notoriously challenging to establish.
Here, we employ quantum weight $K_{xx}$, a direct measure of ground state quantum geometry and topology, for non-metallic materials as developed by Onishi and Fu  \cite{onishi2024quantum}. 
In this quantum weight model tailored for gapped systems, a higher quantum weight indicates less localization of electrons, which leads to stronger topological quantum effects.

The quantum weight $K_{xx}$ is evaluated by modifying the $f$-sum rule weighted by the inverse frequency as follows

\begin{equation}
K_{x x}=\frac{2 \hbar}{e^2} \int_0^{\infty} \frac{\operatorname{Re}\left[\sigma_{x x}(\omega)\right]}{\omega} \mathrm{d} \omega=\frac{2 \hbar}{e^2} \int_0^{\infty} \text{Im}\;[\varepsilon(\omega)] \mathrm{d} \omega
\end{equation}
where $\varepsilon(\omega)$ represents the dimensionless dielectric response function.
In principle, the integral is carried out over an infinite frequency range. Specifically, in this work, we use a cutoff frequency of $\omega_{\text{max}}=50$ eV, and assume $\varepsilon(\omega)$ vanishes for larger frequencies to follow Hung et al \cite{hung2024universal}. Based on density functional theory (DFT) results, Hung et al. \cite{hung2024universal} have already trained an E(3)-equivariant graph neural network to predict optical properties, including the dielectric function with high accuracy.
Thus, for screening purposes, we use their pre-trained model to predict the dielectric function and calculate the quantum weight.

%% file: sections/main_text/m_06_acknowledge.tex
R.O. acknowledges the support from the US Department of Energy (DOE), Office of Science (SC), Basic Energy Sciences (BES), Award No. DE-SC0021940, and Heiwa Nakajima Foundation. Z.W. thanks the support from National Science Foundation (NSF) ITE-2345084. A.C. and M.C. thank the support from NSF Designing Materials to Revolutionize and Engineer our Future (DMREF) Program with Award No. DMR-2118448. M.L. acknowledges the support from the Class of 1947 Career Development Chair and the support from R. Wachnik.\\

%% file: sections/si_text/si_dataset.tex
\section{Text-mined database of inorganic crystal synthesis}

\begin{figure}[H]
\includegraphics[width=0.9\textwidth]{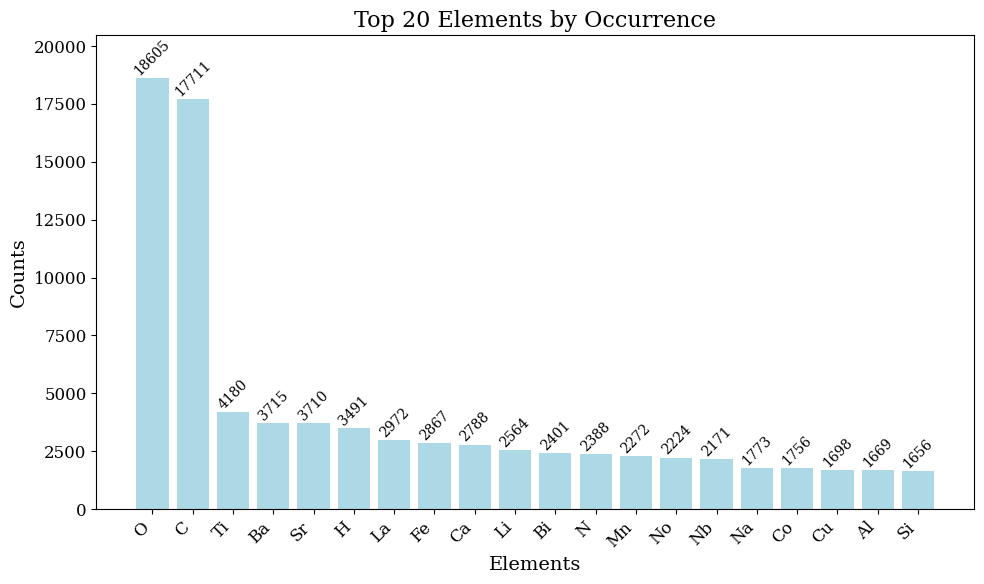}
\caption{\textbf{Occurrences of elements in the 19488 chemical equation training data points.}   
The top 20 elements with highest occurence frequencies are plotted in the histogram.}
\label{fig_si_elems}
\end{figure}

This study utilized the CEDER database as a text-mined dataset of inorganic material synthesis recipes \cite{kononova2019text}. The database contains 19,488 synthesis records extracted from 53,538 paragraphs of scientific literature using natural language processing (NLP) techniques. Figure \ref{fig_si_elems} presents the occurrence of elements in the CEDER database, supporting that the dataset covers 20 chemical elements in the chemical equations for inorganic crystal synthesis.

\begin{figure}[H]
\includegraphics[width=0.9\textwidth]{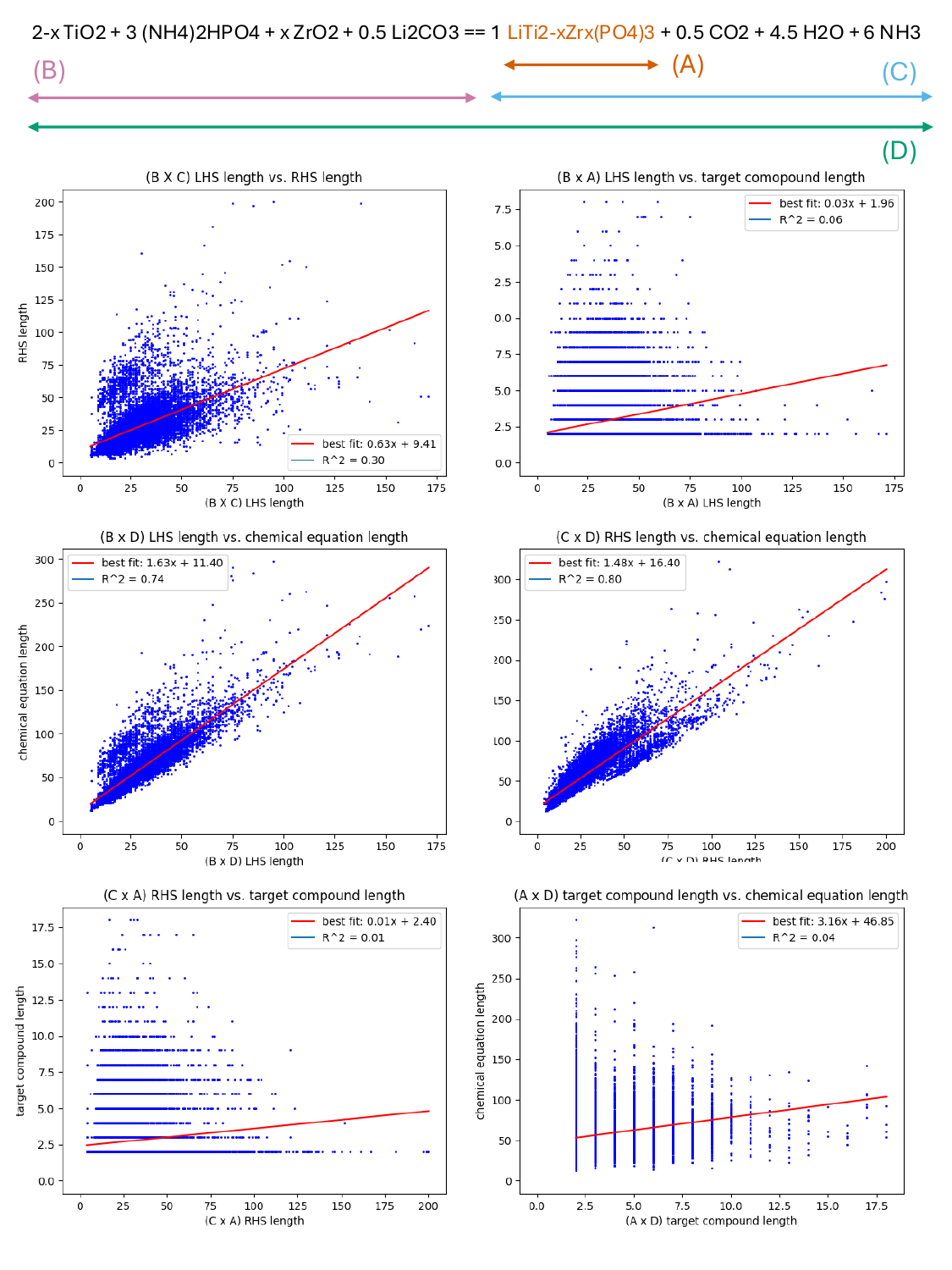}
\caption{\textbf{Relationship between character counts in chemical formulas.}
The scatter plots depict the correlations between the lengths of characters in different components of chemical equations: the left-hand side (LHS), right-hand side (RHS), target compound (TGT), and the complete chemical equation (CEQ). Each plot also shows the best linear fit (in red).}
\label{fig_si_length}
\end{figure}

Figure \ref{fig_si_length} shows the relationship of the lengths of the strings organizing the chemical equations. It suggests the lengths of the left-hand sides (LHS), right-hand sides (RHS), and the total chemical equations (CEQ) have positive correlations, which supports the principle that the chemical equations need to be balanced so that all the elements are preserved through the chemical reaction. Target compounds, on the other hand, have a lower correlation with the lengths of the whole chemical equations.

%% file: sections/si_text/si_training.tex
\section{Fine-tuning of the LLM models}

\begin{figure}[H]
\includegraphics[width=\textwidth]{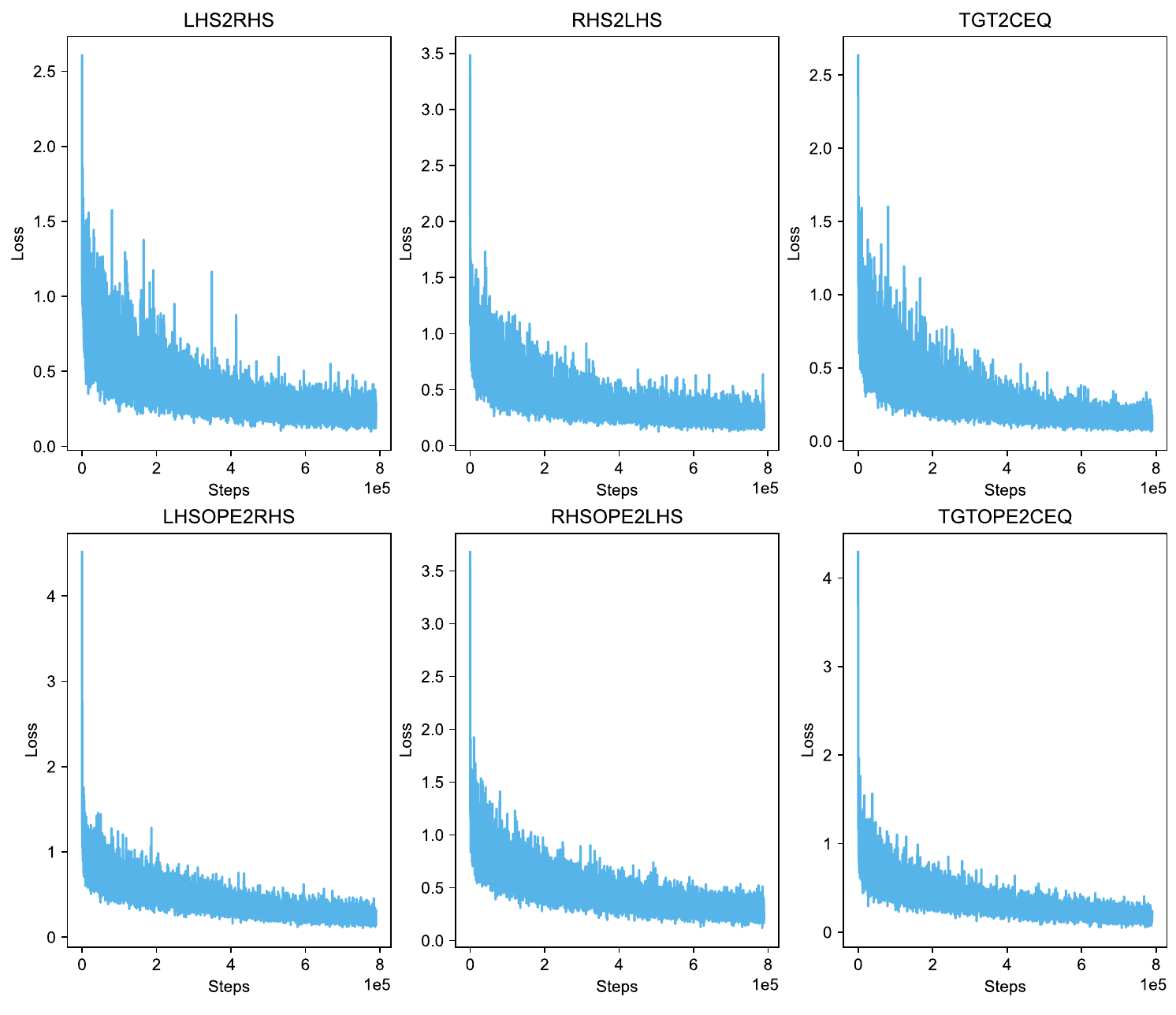}
\caption{\textbf{Training loss curve for multiple LLM prediction tasks for synthesis pathways.} 
We plot the training loss vs. iterated data points for tasks LHS2RHS, RHS2LHS, TGT2CEQ, LHSOPE2RHS RHSOPE2LHS, and TGTOPE2CEQ, respectively.
}
\label{fig_si_loss}
\end{figure}

%% file: sections/si_text/si_inference_tab.tex
\newpage
\section{Inference results}

We present the inference results of text generation using the fine-tuned models. The results include the predictions for LHS2RHS, RHS2LHS, TGT2CEQ, LHSOPE2RHS, RHSOPE2LHS, and TGTOPE2CEQ models. The tables show the top 30 most accurate outcomes for each model based on the prediction similarity metrics. For benchmarking purposes, we present ground-truth text, equations generated by the fine-tuned models, and those from the pre-trained models without fine-tuning. The letters in blue color show the text strings from LLM. For each prediction, we show the predicted accuracy using generalized Tanimoto similarity (GTS) and Jaccard Similarity (JS). 

For each model, only alphabets, numbers, and a few special symbols were accepted. Any unrecognized symbols were eliminated. To improve the appearance of the equations, we replaced `$->$` with `$\rightarrow$` for LHS2RHS (and LHSOPE2RHS) predictions, and `$<-$` with `$\leftarrow$` for RHS2LHS (and RHSOPE2LHS) predictions.

\newpage
\subsubsection*{Table for LHS2RHS}

The LHS2RHS model predicts the right-hand side (RHS) of a chemical equation given the left-hand side (LHS) as the input. The objective of this model is to infer the likely products from a given set of reactants. The data representation for this model involves reactants followed by a separator symbol (`$\rightarrow$'), with the model generating the expected products.

\input{sections/si_text/si_tab_lhs2rhs}

\newpage
\subsubsection*{Table for RHS2LHS}

The RHS2LHS model performs the reverse operation of LHS2RHS, where the model is given the products on the right-hand side (RHS) of a chemical equation and is tasked with predicting the reactants on the left-hand side (LHS). This model helps determine the necessary reactants to produce a target product. The data representation for this model involves reactants followed by a separator symbol (`$\leftarrow$`), with the model generating the expected products.

\input{sections/si_text/si_tab_rhs2lhs}

\newpage
\subsubsection*{Table for TGT2CEQ}

The TGT2CEQ model generates the entire chemical equation when given only the target compound as input. The model predicts both the reactants (LHS) and products (RHS) necessary to synthesize the target compound. If the model produces no text output for a given prediction, we print `Error` for easier accuracy calculation and handling.

\input{sections/si_text/si_tab_tgt2ceq}

\newpage
\subsubsection*{Table for LHSOPE2RHS}

The LHSOPE2RHS model is similar to LHS2RHS but includes synthesis operations (OPE) as part of the input. The model receives the left-hand side (LHS) reactants along with additional operation details (such as heating, quenching, mixing, etc.), and predicts the final products (RHS) after those operations.

\input{sections/si_text/si_tab_lhsope2rhs}

\newpage
\subsubsection*{Table for RHSOPE2LHS}

The RHSOPE2LHS model predicts the left-hand side (LHS) reactants from the products (RHS) and additional synthesis operation information. This model is robust against adding information related to the operation steps needed to produce a specific target compound using operations like heating, quenching, or mixing.

\input{sections/si_text/si_tab_rhsope2lhs}

\newpage
\subsubsection*{Table for TGTOPE2CEQ}

The TGTOPE2CEQ model extends the TGT2CEQ task by incorporating synthesis operations along with the target compound. The model is given the target compound and additional operation details, and it generates both the left-hand side (LHS) and right-hand side (RHS) of the chemical equation. If the model fails to generate any output, `Error` is printed for easier tracking of prediction accuracy.

\input{sections/si_text/si_tab_tgtope2ceq}

%% file: sections/si_text/si_tab_lhs2rhs.tex
\begin{longtable}{c|c|p{10cm}|c|c}
\caption{Selected examples of LHS2RHS}\label{tab_s_lhs2rhs}\\
No. &  Model & Text & GTS & JS \\
\Xhline{5\arrayrulewidth}
\endfirsthead
\multicolumn{4}{c}{{\tablename\ \thetable{} -- continued from previous page}} \\
No. &  Model & Text & GTS & JS \\
\hline
\endhead
\hline
\multicolumn{4}{r}{{Continued on next page}} \\
\endfoot
\hline
\endlastfoot
1 & Ground truth & 0.1 CuO + 1 Fe2O3 + 0.32 NiO + 0.58 ZnO $\rightarrow$ 1 Ni0.32Zn0.58Cu0.1Fe2O4 & 1.0 & 1.0 \\
\hline
1 & w/ fine-tuning & 0.1 CuO + 1 Fe2O3 + 0.32 NiO + 0.58 ZnO $\rightarrow$  \textcolor{blue}{1 Ni0.32Zn0.58Cu0.1Fe2O4} & 1.0 & 1.0 \\
\hline
1 & w/o fine-tuning & 0.1 CuO + 1 Fe2O3 + 0.32 NiO + 0.58 ZnO $\rightarrow$  \textcolor{blue}{0.38 NiO $\rightarrow$ 0.38 NiO $\rightarrow$} & 0.238 & 0.0 \\
\Xhline{5\arrayrulewidth}
2 & Ground truth & 0.015 Nd2O3 + 0.835 Fe2O3 + 0.15 Cr2O3 + 1 CuO $\rightarrow$ 1 CuCr0.3R0.03Fe1.67O4 & 1.0 & 1.0 \\
\hline
2 & w/ fine-tuning & 0.015 Nd2O3 + 0.835 Fe2O3 + 0.15 Cr2O3 + 1 CuO $\rightarrow$  \textcolor{blue}{1 CuCr0.3R0.03Fe1.67O4} & 1.0 & 1.0 \\
\hline
2 & w/o fine-tuning & 0.015 Nd2O3 + 0.835 Fe2O3 + 0.15 Cr2O3 + 1 CuO $\rightarrow$  \textcolor{blue}{0.038 Cr2O3 + 1 CuO $\rightarrow$} & 0.466 & 0.143 \\
\Xhline{5\arrayrulewidth}
3 & Ground truth & 1 Se + 1.94 Cu + 0.02 Ga $\rightarrow$ 1 Cu1.94A0.02Se & 1.0 & 1.0 \\
\hline
3 & w/ fine-tuning & 1 Se + 1.94 Cu + 0.02 Ga $\rightarrow$  \textcolor{blue}{1 Cu1.94A0.02Se} & 1.0 & 1.0 \\
\end{longtable}

%% file: sections/si_text/si_tab_rhs2lhs.tex
\begin{longtable}{c|c|p{10cm}|c|c}
\caption{Selected examples of RHS2LHS}\label{tab_s_rhs2lhs}\\
No. &  Model & Text & GTS & JS \\
\Xhline{5\arrayrulewidth}
\endfirsthead
\multicolumn{4}{c}{{\tablename\ \thetable{} -- continued from previous page}} \\
No. &  Model & Text & GTS & JS \\
\hline
\endhead
\hline
\multicolumn{4}{r}{{Continued on next page}} \\
\endfoot
\hline
\endlastfoot
1 & Ground truth & 1 NaCaAlF6 $\leftarrow$ 1 CaF2 + 1 AlF3 + 1 NaF & 1.0 & 1.0 \\
\hline
1 & w/ fine-tuning & 1 NaCaAlF6 $\leftarrow$  \textcolor{blue}{1 AlF3 + 1 CaF2 + 1 NaF} & 1.0 & 1.0 \\
\hline
1 & w/o fine-tuning & 1 NaCaAlF6 $\leftarrow$  \textcolor{blue}{NaCaAlF6 $\leftarrow$ NaCaAlF6 $\leftarrow$} & 0.422 & 0.0 \\
\Xhline{5\arrayrulewidth}
2 & Ground truth & 1 (1-x)(Na0.5K0.5)NbO3-xLiNbO3 + 0.5 CO2 $\leftarrow$ 0.25-0.25$*$x Na2CO3 + 0.25-0.25$*$x K2CO3 + 0.5 Nb2O5 + 0.5$*$x Li2CO3 & 1.0 & 1.0 \\
\hline
2 & w/ fine-tuning & 1 (1-x)(Na0.5K0.5)NbO3-xLiNbO3 + 0.5 CO2 $\leftarrow$  \textcolor{blue}{0.25-0.25$*$x Na2CO3 + 0.25-0.25$*$x K2CO3 + 0.5 Nb2O5 + 0.5$*$x Li2CO3} & 1.0 & 1.0 \\
\hline
2 & w/o fine-tuning & 1 (1-x)(Na0.5K0.5)NbO3-xLiNbO3 + 0.5 CO2 $\leftarrow$  \textcolor{blue}{0.5 CO2 $\leftarrow$ 0.5 CO2 $\leftarrow$ 0.5 CO2 $\leftarrow$ 0.5 CO2 $\leftarrow$ 0.5 CO2 $\leftarrow$ 0.5 CO2 $\leftarrow$} & 0.365 & 0.1 \\
\Xhline{5\arrayrulewidth}
3 & Ground truth & 1 CaCu3Ti4O12 + 1 CO2 $\leftarrow$ 3 CuO + 4 TiO2 + 1 CaCO3 & 1.0 & 1.0 \\
\hline
3 & w/ fine-tuning & 1 CaCu3Ti4O12 + 1 CO2 $\leftarrow$  \textcolor{blue}{3 CuO + 4 TiO2 + 1 CaCO3} & 1.0 & 1.0 \\
\end{longtable}

%% file: sections/si_text/si_tab_tgt2ceq.tex
\begin{longtable}{c|c|p{10cm}|c|c}
\caption{Selected examples of TGT2CEQ}\label{tab_s_tgt2ceq}\\
No. &  Model & Text & GTS & JS \\
\Xhline{5\arrayrulewidth}
\endfirsthead
\multicolumn{4}{c}{{\tablename\ \thetable{} -- continued from previous page}} \\
No. &  Model & Text & GTS & JS \\
\hline
\endhead
\hline
\multicolumn{4}{r}{{Continued on next page}} \\
\endfoot
\hline
\endlastfoot
1 & Ground truth & Fe2Ni0.32Cu0.1Zn0.58O4 $||$ 0.1 CuO + 1 Fe2O3 + 0.32 NiO + 0.58 ZnO $\rightarrow$ 1 Ni0.32Zn0.58Cu0.1Fe2O4 & 1.0 & 1.0 \\
\hline
1 & w/ fine-tuning & Fe2Ni0.32Cu0.1Zn0.58O4 $||$ \textcolor{blue}{0.1 CuO + 1 Fe2O3 + 0.32 NiO + 0.58 ZnO $\rightarrow$ 1 Ni0.32Cu0.1Zn0.58Fe2O4} & 1.0 & 0.833 \\
\hline
1 & w/o fine-tuning & Fe2Ni0.32Cu0.1Zn0.58O4 $||$  \textcolor{blue}{0.01} & 0.0 & 0.0 \\
\Xhline{5\arrayrulewidth}
2 & Ground truth & Nd0.03Cr0.3Fe1.67CuO4 $||$ 0.015 Nd2O3 + 0.835 Fe2O3 + 0.15 Cr2O3 + 1 CuO $\rightarrow$ 1 CuCr0.3R0.03Fe1.67O4 & 1.0 & 1.0 \\
\hline
2 & w/ fine-tuning & Nd0.03Cr0.3Fe1.67CuO4 $||$ \textcolor{blue}{0.835 Fe2O3 + 0.15 Cr2O3 + 0.015 Nd2O3 + 1 CuO $\rightarrow$ 1 CuCr0.3R0.03Fe1.67O4} & 1.0 & 1.0 \\
\hline
2 & w/o fine-tuning & Nd0.03Cr0.3Fe1.67CuO4 $||$  \textcolor{blue}{0.0.0.0.0.0.0.0.0.0.0.0.0.0.0.0.0.0.0.0.0.0.0.0.0.0.0.0.0.0.0.0.0.0.0.0.} & 0.0 & 0.0 \\
\Xhline{5\arrayrulewidth}
3 & Ground truth & Cu1.94Ga0.02Se $||$ 1 Se + 1.94 Cu + 0.02 Ga $\rightarrow$ 1 Cu1.94A0.02Se & 1.0 & 1.0 \\
\hline
3 & w/ fine-tuning & Cu1.94Ga0.02Se $||$ \textcolor{blue}{1 Se + 1.94 Cu + 0.02 Ga $\rightarrow$ 1 Cu1.94A0.02Se} & 1.0 & 1.0 \\
\end{longtable}

%% file: sections/si_text/si_tab_lhsope2rhs.tex
\begin{longtable}{c|c|p{10cm}|c|c}
\caption{Selected examples of LHSOPE2RHS}\label{tab_s_lhsope2rhs}\\
No. &  Model & Text & GTS & JS \\
\Xhline{5\arrayrulewidth}
\endfirsthead
\multicolumn{4}{c}{{\tablename\ \thetable{} -- continued from previous page}} \\
No. &  Model & Text & GTS & JS \\
\hline
\endhead
\hline
\multicolumn{4}{r}{{Continued on next page}} \\
\endfoot
\hline
\endlastfoot
1 & Ground truth & 0.015 Nd2O3 + 0.835 Fe2O3 + 0.15 Cr2O3 + 1 CuO : Mix Shape Heat Quench Heat Heat Heat $\rightarrow$ 1 CuCr0.3R0.03Fe1.67O4 & 1.0 & 1.0 \\
\hline
1 & w/ fine-tuning & 0.015 Nd2O3 + 0.835 Fe2O3 + 0.15 Cr2O3 + 1 CuO : Mix Shape Heat Quench Heat Heat Heat $\rightarrow$  \textcolor{blue}{1 CuCr0.3R0.03Fe1.67O4} & 1.0 & 1.0 \\
\hline
1 & w/o fine-tuning & 0.015 Nd2O3 + 0.835 Fe2O3 + 0.15 Cr2O3 + 1 CuO : Mix Shape Heat Quench Heat Heat Heat $\rightarrow$  \textcolor{blue}{Q CuO : Mix Shape Heat} & 0.239 & 0.0 \\
\Xhline{5\arrayrulewidth}
2 & Ground truth & 0.1 CuO + 1 Fe2O3 + 0.32 NiO + 0.58 ZnO : Mix Heat Shape Shape Heat $\rightarrow$ 1 Ni0.32Zn0.58Cu0.1Fe2O4 & 1.0 & 1.0 \\
\hline
2 & w/ fine-tuning & 0.1 CuO + 1 Fe2O3 + 0.32 NiO + 0.58 ZnO : Mix Heat Shape Shape Heat $\rightarrow$  \textcolor{blue}{1 Ni0.32Zn0.58Cu0.1Fe2O4} & 1.0 & 1.0 \\
\hline
2 & w/o fine-tuning & 0.1 CuO + 1 Fe2O3 + 0.32 NiO + 0.58 ZnO : Mix Heat Shape Shape Heat $\rightarrow$  \textcolor{blue}{Heat Shape Heat $\rightarrow$ Heat} & 0.0 & 0.0 \\
\Xhline{5\arrayrulewidth}
3 & Ground truth & 0.5 SnO2 + 0.05 Gd2O3 + 0.9 BaCO3 + 0.25 In2O3 : Mix Shape Heat $\rightarrow$ 1 Ba0.9Gd0.1Sn0.5In0.5O2.8 + 0.9 CO2 & 1.0 & 1.0 \\
\hline
3 & w/ fine-tuning & 0.5 SnO2 + 0.05 Gd2O3 + 0.9 BaCO3 + 0.25 In2O3 : Mix Shape Heat $\rightarrow$  \textcolor{blue}{1 Ba0.9Gd0.1Sn0.5In0.5O2.8 + 0.9 CO2} & 1.0 & 1.0 \\
\end{longtable}

%% file: sections/si_text/si_tab_rhsope2lhs.tex
\begin{longtable}{c|c|p{10cm}|c|c}
\caption{Selected examples of RHSOPE2LHS}\label{tab_s_rhsope2lhs}\\
No. &  Model & Text & GTS & JS \\
\Xhline{5\arrayrulewidth}
\endfirsthead
\multicolumn{4}{c}{{\tablename\ \thetable{} -- continued from previous page}} \\
No. &  Model & Text & GTS & JS \\
\hline
\endhead
\hline
\multicolumn{4}{r}{{Continued on next page}} \\
\endfoot
\hline
\endlastfoot
1 & Ground truth & 1 40BaO-20TiO2-40SiO2 + 40 CO2: Mix Heat Shape Heat Heat $\leftarrow$ 40 SiO2 + 20 TiO2 + 40 BaCO3 & 1.0 & 1.0 \\
\hline
1 & w/ fine-tuning & 1 40BaO-20TiO2-40SiO2 + 40 CO2: Mix Heat Shape Heat Heat $\leftarrow$  \textcolor{blue}{40 SiO2 + 20 TiO2 + 40 BaCO3} & 1.0 & 1.0 \\
\hline
1 & w/o fine-tuning & 1 40BaO-20TiO2-40SiO2 + 40 CO2: Mix Heat Shape Heat Heat $\leftarrow$  \textcolor{blue}{Heat Shape Heat Heat $\leftarrow$ Heat} & 0.0 & 0.0 \\
\Xhline{5\arrayrulewidth}
2 & Ground truth & 1 (1-x)(Na0.5K0.5)NbO3-xLiNbO3 + 0.5 CO2: Mix Dry Heat Mix Dry Shape Heat Mix Heat $\leftarrow$ 0.25-0.25$*$x Na2CO3 + 0.25-0.25$*$x K2CO3 + 0.5 Nb2O5 + 0.5$*$x Li2CO3 & 1.0 & 1.0 \\
\hline
2 & w/ fine-tuning & 1 (1-x)(Na0.5K0.5)NbO3-xLiNbO3 + 0.5 CO2: Mix Dry Heat Mix Dry Shape Heat Mix Heat $\leftarrow$  \textcolor{blue}{0.25-0.25$*$x Na2CO3 + 0.25-0.25$*$x K2CO3 + 0.5 Nb2O5 + 0.5$*$x Li2CO3} & 1.0 & 1.0 \\
\hline
2 & w/o fine-tuning & 1 (1-x)(Na0.5K0.5)NbO3-xLiNbO3 + 0.5 CO2: Mix Dry Heat Mix Dry Shape Heat Mix Heat $\leftarrow$  \textcolor{blue}{(1-x)(Na0.5K0.5)NbO3-xLiNbO3 + 0.5 CO2: Mix Dry Shape Heat Mix He} & 0.439 & 0.133 \\
\Xhline{5\arrayrulewidth}
3 & Ground truth & 1 CaCu3Ti4O12 + 1 CO2: LiquidGrind Mix Shape Heat Heat $\leftarrow$ 3 CuO + 4 TiO2 + 1 CaCO3 & 1.0 & 1.0 \\
\hline
3 & w/ fine-tuning & 1 CaCu3Ti4O12 + 1 CO2: LiquidGrind Mix Shape Heat Heat $\leftarrow$  \textcolor{blue}{3 CuO + 4 TiO2 + 1 CaCO3} & 1.0 & 1.0 \\
\end{longtable}

%% file: sections/si_text/si_tab_tgtope2ceq.tex
\begin{longtable}{c|c|p{10cm}|c|c}
\caption{Selected examples of TGTOPE2CEQ}\label{tab_s_tgtope2ceq}\\
No. &  Model & Text & GTS & JS \\
\Xhline{5\arrayrulewidth}
\endfirsthead
\multicolumn{4}{c}{{\tablename\ \thetable{} -- continued from previous page}} \\
No. &  Model & Text & GTS & JS \\
\hline
\endhead
\hline
\multicolumn{4}{r}{{Continued on next page}} \\
\endfoot
\hline
\endlastfoot
1 & Ground truth & Cu1.94Ga0.02Se: Heat LiquidGrind Heat $||$ 1 Se + 1.94 Cu + 0.02 Ga $\rightarrow$ 1 Cu1.94A0.02Se & 1.0 & 1.0 \\
\hline
1 & w/ fine-tuning & Cu1.94Ga0.02Se: Heat LiquidGrind Heat $||$ \textcolor{blue}{1 Se + 1.94 Cu + 0.02 Ga $\rightarrow$ 1 Cu1.94A0.02Se} & 1.0 & 1.0 \\
\hline
1 & w/o fine-tuning & Cu1.94Ga0.02Se: Heat LiquidGrind Heat $||$  \textcolor{blue}{0.0 0.0 0.0 0.0 0.0 0.0 0.0 0.0 0.0 0.0 0.0} & 0.0 & 0.0 \\
\Xhline{5\arrayrulewidth}
2 & Ground truth & BaCu0.85Oy: Heat Shape Heat Heat Quench $||$ 0.85 CuO + 1 BaCO3 $\rightarrow$ 1 BaCu0.85Oy + 1 CO2 + 0.925-0.5$*$y O2 & 1.0 & 1.0 \\
\hline
2 & w/ fine-tuning & BaCu0.85Oy: Heat Shape Heat Heat Quench $||$ \textcolor{blue}{0.85 CuO + 1 BaCO3 $\rightarrow$ 1 BaCu0.85Oy + 1 CO2 + 0.425-0.5$*$y O2} & 1.0 & 0.818 \\
\hline
2 & w/o fine-tuning & BaCu0.85Oy: Heat Shape Heat Heat Quench $||$  \textcolor{blue}{1} & 0.0 & 0.1 \\
\Xhline{5\arrayrulewidth}
3 & Ground truth & Co20Cu80: $||$ 80 Cu + 20 Co $\rightarrow$ 1 Co20Cu80 & 1.0 & 1.0 \\
\hline
3 & w/ fine-tuning & Co20Cu80: $||$ \textcolor{blue}{80 Cu + 20 Co $\rightarrow$ 1 Cu80Co20} & 1.0 & 0.778 \\
\end{longtable}